\documentclass[pdftex,twocolumn,epjc3]{svjour3}          

\RequirePackage[T1]{fontenc}

\smartqed  

\RequirePackage{graphicx}
\RequirePackage{mathptmx}      
\RequirePackage{flushend}
\RequirePackage[numbers,sort&compress]{natbib}
\RequirePackage[colorlinks,citecolor=blue,urlcolor=blue,linkcolor=blue]{hyperref}
\usepackage{amssymb}
\usepackage{amsmath}

\journalname{Eur. Phys. J. C}

\begin{document}

\title{Mimetic gravity: mimicking the dynamics of the primeval universe in the context of loop quantum cosmology}


\author{Eunice Bezerra \thanksref{e1}
        \and
        Oswaldo D. Miranda \thanksref{e2} 
}

\thankstext{e1}{e-mail: eunicebezerra1@gmail.com}
\thankstext{e2}{e-mail: oswaldo.miranda@inpe.br}

\institute{Divis\~ao de Astrof\'isica, Instituto Nacional de Pesquisas Espaciais (INPE), Av. dos Astronautas 1758, S\~ao Jos\'e dos Campos, 12227-010 SP, Brazil
}

\date{Received: date / Accepted: date}

\maketitle

\begin{abstract}
Mimetic gravity can be described as a formulation capable of mimicking different evolutionary scenarios regarding the universe dynamics. Notwithstanding its initial aim of producing a similar evolution to the one expected from the dark components of the standard cosmology, a recent association with loop quantum cosmology could also provide interesting results. In this work, we reinterpret the physics behind the curvature potential of mimetic gravity description of loop quantum cosmology. Furthermore, we also test the compatibility of our formulation for a Higgs-type field, proving that the mimetic curvature potential can mimic the dynamics from a Higgs inflationary model. Additionally, we discuss possible scenarios that emerge from the relationship between matter and mimetic curvature and, within certain limits, describe results for the primeval universe dynamics obtained by other authors.
\end{abstract}

\section{Introduction}

Loop quantum gravity (LQG) is an attempt to quantize gravity by performing a nonperturbative quantization of general relativity (GR) \cite{martineau/17} at kinematic level that has been showing progress during the last few years. Mainly, its cosmological description called loop quantum cosmology (LQC), see \cite{ashtekar/11, pullin/11, agullo/16} for a dedicated review. LQC overcomes the kinematic character of LQG through cosmological dynamics. Moreover, it naturally solves the initial singularity problem by replacing it with a bounce for, at least, the most common cosmological models \cite{agullo/17}. 

Effective LQC is a compelling proposal because it results in regular solutions. It does not matter if we are analyzing the primordial universe evolution from matter, curvature or scale factor angle, the solutions do not diverge \cite{ashtekar/112}. In order to reproduce LQC results, many approaches have been tested, including the ones with massless fields, different potential shapes and so on. Among them, a recent work from Langlois \textit{et al.} \cite{langlois/17} showed how to recover the Effective LQC dynamics through a class of scalar-tensor theories, in which the Mimetic Gravity (MG) of Chamseddine and Mukhanov was included (see, in particular, \cite{chamseddine/17, chamseddine/172}). The remarkable feature in this description is the treatment of curved space-times. The strategy employed to incorporate curvature provided a new window we intend to explore.

The MG description of LQC dynamics can be interpreted in such a way that enables us to follow the evolution of a scalar field whose potential is intrinsically coupled to a curved background. Because its nature and the fact that the Higgs field is the only scalar field currently observed, a Higgs-type field is a perfect candidate to test our approach. Furthermore, the possibility of relating the mimetic field to the Higgs mechanism presented in \cite{chamseddine/182} and \cite{chamseddine/18} fortify our idea.

In this work, we aim to emphasize how powerful and versatile the mimetic formulation of LQC is. First, the MG curvature potential is interpreted as the geometric response to the presence of matter onto spacetime, which allows the study of the universe evolution without considering the field potential directly. Next, we analyze the implications regarding the different interpretations of the curvature role in the universe dynamics. Furthermore, once the general solution for the Hubble parameter is obtained, we will show that the dynamics of the universe during the inflationary period can be described within the framework of the mimetic representation of LQC. Although it is possible through MG to mimic any scalar field, here we will show that the potential for curvature in the mimetic description of LQC can produce the same evolution for the Hubble parameter as that derived from the scenario known as Higgs Inflation (HI).

The paper is organized as follows. In Sect. \ref{overview}, we summarize only the essential aspects of LQC and MG, emphasizing the universe evolution during the early times. In Sect. \ref{discussion}, we expose our interpretation of the result presented in Langlois \textit{et al.} \cite{langlois/17} and how we construct an alternative evolutionary scenario by applying our formalism. We also show how the mimetic curvature potential must behave to reproduce the HI dynamics. Therefore, from the viewpoint of the dynamics described by the Hubble parameter, the inflationary Higgs phase can be perfectly mimicked by the curvature potential introduced by Langlois \textit{et al.} \cite{langlois/17}. Besides, we use this potential to adjust the HI and effective mimetic LQC energy scales, displaying the compatibility between them. We conclude Sect. \ref{discussion} with two Sects. (\ref{case_1} and \ref{case_2}) that discuss possible interpretations of the curvature potential from MG representation of LQC. In fact, we have shown that it is possible to recover, within certain limits, previous analyzes of \cite{sadjadi/13} (for $k = 0$) or that studied in \cite{mielczarek/092} (for $k = 1$). Finally, we highlight the most relevant implications of our proposal in Sect. \ref{final}. 
\section{Overview}
\label{overview}

The following computations were developed using the natural unity system. Consequently, the velocity of light $c$ and reduced Planck constant $\hbar$ are unitary ($ c = \hbar = 1$). Besides, the Newtonian gravitational constant $G$, Planck length $\ell_{Pl}$, Planck mass $m_{Pl}$ and reduced Planck mass $M_{Pl}$ are related by $m_{Pl} = \ell^{-1}_{Pl} = (\sqrt{G})^{-1} = \sqrt{8\pi} M_{Pl}$.

\subsection{Loop Quantum Cosmology}
\label{slqc}

The key element that makes LQG different from other approaches of quantum gravity is the holonomy introduction. The Ashtekar connection $A^{i}_{a}$ and its conjugated momentum $E^{a}_{i}$ \footnote{Meanwhile the indices a, b, c,... refer to the spatial manifold $\Sigma$, i, j, k,... are internal indices related to the fiducial cell which is defined as a finite region introduced to avoid integrals over the space-time infinite region \cite{langlois/17}.} are the canonical variables of LQG \cite{langlois/17, mielczarek/12} whose forms are given by
\begin{equation}
    A^{i}_{a} = c (dx^{i})_{a} \;\;\; \mathrm{and}  \; \;\; E^{a}_{i} = p \left(\frac{\partial}{\partial x^{i}}\right)^{a},
    \label{ashtekar}
    \end{equation}
where $x^{i}$ refers to space coordinates and $\gamma \simeq 0.2375$ represents the Barbero-Immirzi parameter \cite{meissner/04, mielczarek/102}. The variables $p$ and $c$ are defined with respect to the scale factor $a$ and its time derivative $\dot{a}$ as 
\begin{equation}
p=a^{2}, \; c = \gamma \dot{a}/{N} \;\; \mathrm{and} \;\; \{c, p\}= \frac{8\pi G \gamma}{3},    
\end{equation}
being $N$ the lapse function. However, instead of trying to implement $A^{i}_{a}$ or $c$ as quantum operator, the holonomy (as a function of $A^{i}_{a}$) is the one defined as fundamental operator \cite{langlois/17}, resulting in the so-called holonomy corrections.

LQC incorporates the quantization scheme and techniques from LQG and applies them for homogeneous and isotropic space-times \cite{zhang/07, langlois/17}. Hence, LQG can be considered a canonical quantization of gravity, meanwhile, LQC corresponds to a canonical quantization of homogeneous and isotropic space-times \cite{zhang/07}. In LQC, the holonomy is considered around a loop with square shape due to the symmetries of Friedmann-Robertson-Walker (FRW) space-times. Because $p$ is the variable related to $E^{a}_{i}$, it is the one promoted to area operator, presenting a discrete spectrum \cite{langlois/17}. Thus, there is a minimum area value, usually referred to as $\Delta = 2\sqrt{3}\pi \gamma \ell^{2}_{Pl}$ that limits the size of the loop as a fundamental structure of space-time. The relation between $\Delta$ and the face physical area $|p|$ is described by $\bar{\mu}^{2} = \Delta/{|p|}$ \cite{ashtekar/062}. This procedure determines the loop area, being named $\bar{\mu}$-scheme \cite{mielczarek/12, ashtekar/11}.

In this quantum cosmological scenario, the universe dynamics is determined by the Effective LQC version of the Friedmann equations and continuity equation. The LQC effective Friedmann equation can be obtained from the evolution of the observable $p$, which corresponds to its equation of motion, dictated by
\begin{equation}
\dot{p} = \frac{dp}{dt} = \{p , {\cal{C}}_{H}\} = \left\{p , {\int} d^{3}\vec{x} N {\cal{H}} \right\},
\label{obs}
\end{equation}
where ${\cal{C}}_{H} = \int d^{3}\vec{x} N {\cal{H}}$ is the Hamiltonian constraint. First, the Hamiltonian receives a classical treatment in which it acquires the shape \cite{langlois/17}
\begin{equation}
{\cal{H}}_{\mathrm{eff}} = - \frac{3 p^{3/2}}{8 \pi G\Delta \gamma^{2}} \sin^{2}\bar{\mu}c + \frac{\pi^{2}_{\varphi}}{2p^{3/2}} + p^{3/2}V(\varphi),
\label{h_class}
\end{equation}
with $\pi_{\varphi}$ and $V(\varphi)$ representing the momentum and potential of the matter content defined by the scalar field $\varphi$, respectively. Second, as the Hamiltonian constraint is weakly equal to zero, plugging (\ref{h_class}) into (\ref{obs}) is going to result in
\begin{equation}
\dot{p} = 2N \frac{p}{\gamma \sqrt{\Delta}} \sin \bar{\mu}c \cos \bar{\mu}c, \;\; \mathrm{where} \;\;\sin^{2} \bar{\mu}c = \frac{\rho}{\rho_{c}}.
\label{evop}
\end{equation}
At this point, it is clear how the sine function restricts the relation between the matter energy density $\rho$ and its critical value $\rho_{c}$ to the range $0 \lesssim \rho/ \rho_{c} \leq 1$ \cite{bojowald/08}. Here, $\rho$ presents the classical form
\begin{equation}
\rho = \frac{\pi^{2}_{\varphi}}{2p^{3}} + V(\varphi),
\label{rho2}
\end{equation}
and 
\begin{equation}
\rho_{c} =  \frac{3}{8\pi G \gamma^{2}\Delta} \simeq 0.82 \rho_{Pl},
\label{rhoc}
\end{equation}
where $\rho_{Pl}$ is the matter energy density at Planck scale. Finally, from (\ref{evop}), it is straight to obtain the Hubble parameter 
\begin{equation}
H^{2} = \left(\frac{\dot{p}}{2Np}\right)^{2} = \frac{8\pi G}{3} \rho \left(1 - \frac{\rho}{\rho_{c}}\right).
\label{h_lqc}
\end{equation}

The equation (\ref{h_lqc}) provides strong implications regarding the LQC evolutionary scenario for the very early times. As we regress in the universe history, the matter energy density is growing more and more. In the Hot Big Bang classical solution,  $\rho$ goes to infinite due to its inverse relationship with time, resulting in the so-called initial singularity. However, from the LQC perspective, the universe undergoes a bounce phase in which the matter content is compressed until $\rho$ achieves a value close to $\rho_{Pl}$ \cite{ashtekar/062}. This can be directly observed from equation (\ref{h_lqc}), where the minimum point of the function $H$ ($H = 0$) occurs for $\rho = \rho_{c}$. Thereupon, the relation $\rho / \rho_{c} = 1$ defines the turn point in scale factor evolution ($\dot{a} = 0$) that means a change in the evolution of universe itself. Consequently, instead of a singularity characterized by an infinite energy density, in LQC, there is a big bounce when the energy density achieves a range close to Planck scale determined by $\rho_{c}$ \cite{ashtekar/062, bojowald/08, sadjadi/13}. 

Another key point to remember is that all physical fields are considered regular at LQC bounce for strong curvature singularities in FRW models. As a result, any matter field used in LQC context must obey the usual expressions for the state equation (EoS)
\begin{equation}
P = w\rho,
\label{state}
\end{equation}
and Klein Gordon equation
\begin{equation}
\ddot{\varphi} + 3H \dot{\varphi} + \frac{d V(\varphi)}{d \varphi} = 0,
\label{klein}
\end{equation}
where $P$ is the matter pressure and $w$ represents the state parameter \cite{ashtekar/112, sadjadi/13}. Moreover, regardless of theory, the continuity equation must be satisfied, which means the matter energy density obeys the expression
\begin{equation}
\dot{\rho} + 3H (P + \rho) = 0.
\label{continuity}
\end{equation}

After the bounce, the standard LQC universe undergoes a period called super-inflation \cite{ashtekar/11, ashtekar/112}. During this phase, the Hubble parameter is extremely dynamical, going from $H = 0$ to its maximum value 
\begin{equation}
H^{2} = \frac{\rho_{c}}{12M^{2}_{Pl}},
\label{hmax}
\end{equation}
when $\rho = (1/2)\rho_{c}$. Nevertheless, this phase should not last long (less than a Planck second) in order to avoid significant changes in the evolution of $a$ and $\varphi$. Furthermore, this rapid growth implies a large friction term in  (\ref{klein}) which takes time to slow down until a point the potential energy is capable of dominating the universe dynamic and producing a slow-roll-type evolution \cite{ashtekar/112}. 

In principle, a bounce phase could enable the field potential to climb the potential well \cite{mielczarek/092}, which would naturally provide the initial condition for the field starts to roll down as expected from the standard inflationary scenario. Indeed, the super-inflation should end with the universe in a suitable state for the beginning of inflation. However, the ratio regarding the kinetic and potential energies seems to determine the qualitative features of dynamical evolution when the initial data is established. Depending on how greater the kinetic energy is regarding the potential energy, the sorter the super-inflation phase is going to be. Moreover, producing an inflationary period with a nearly constant Hubble parameter, in agreement with the full LQC statement $\dot{H} = 0$ at the end of super-inflation, requires specific adjustments for a bounce with the kinetic energy density as the dominant component \cite{ashtekar/112} (a common assumption found in LQC literature). 

The standard Effective LQC Hamiltonian constraint (see equation (\ref{h_class})) implies that quantum gravitational effects are negligible for values of $\rho$ much smaller than $\rho_{Pl}$ \cite{ashtekar/062}, which enables to recover the standard Friedmann equation for flat FRW space-time
\begin{equation}
H^{2} = \frac{\rho}{3M^{2}_{Pl}}.
\label{hend}
\end{equation}
Once both (\ref{h_lqc}) and (\ref{continuity}) are already determined, it is straight to obtain the acceleration equation for standard LQC by computing the time derivative of (\ref{h_lqc}) and summing with (\ref{h_lqc}) itself, obtaining the expression 

\begin{equation}
\dot{H} + H^{2} = \frac{4\pi G}{3}\rho\left[ - 3w -1 + \frac{2\rho}{\rho_{c}} (2 + 3w) \right].
\label{acelqc}
\end{equation}

In summary, from Effective LQC approach, the effective Hamiltonian results in a modified Friedmann equation that only differs from the classical one by a quadratic term of the energy density $\rho^{2}$, besides the universe underwent a bounce phase followed by an inflationary period regardless the matter content assumed. Notwithstanding, between these two stages, a super-inflation period is expected to take place \cite{zhang/07}. During super-inflation, the universe is in a super-accelerated stage $\dot{H} > 0$, meanwhile, along inflationary epoch, $H$ obeys the relation $\dot{H} < 0$ \cite{sadjadi/13}. In this scenario, the gravity presents a repulsive behavior in the deep Planck regime due to quantum geometry \cite{ashtekar/062}, whose effects are negligible for sufficiently small values of $\rho$ ($\rho / \rho_{c} \rightarrow 0$), recovering the classical dynamics of standard cosmology.


\subsection{Mimetic gravity description for loop quantum cosmology}

The mimetic gravity provides a unified geometric description of the universe evolution without any extra dark component \cite{sebastiani/17}. Despite being recently proposed by Chamseddine and Mukhanov in \cite{chamseddine/13} as a way to simulate the dark matter behavior, MG can also overcome cosmological singularities issues through the limiting curvature concept \cite{chamseddine/17, chamseddine/172}. Furthermore, the mimetic representation has been extended to reproduce a plethora of different frameworks (see \cite{sebastiani/17, dutta/18, ganz/18} for further discussions). 

Basically, the MG formulation was built under the concept of disformal transformations as a consequence of GR invariance under diffeomorphism transformations \cite{bekenstein/93, rabochaya/16}. This kind of transformation enables to parameterize $g_{\mu\nu}$ as a function of an auxiliary metric $\tilde{g}_{\mu\nu}$ and a scalar field $\varphi$, the mimetic field, like  \cite{chamseddine/14, sebastiani/17}
\begin{equation}
g_{\mu\nu} = - (\tilde{g}^{\alpha \beta} \partial_{\alpha} \varphi \partial_{\beta} \varphi) \tilde{g}_{\mu\nu}.
\label{maux}
\end{equation}
From (\ref{maux}), two fundamental features emerged. First, the invariance of $g^{\mu\nu}$ under a conformal transformation of $\tilde{g}^{\mu\nu}$ like $\tilde{g}^{\mu\nu} \rightarrow \Omega (t, \textbf{x})^{2}\tilde{g}^{\mu\nu}$. And second, the consistence condition
\begin{equation}
  g^{\mu\nu} \partial_{\mu} \varphi \partial_{\nu} \varphi = -1       
\label{mgconsist}
 \end{equation}
that $\varphi$ must satisfy. These properties can be directly related to the two equivalent formulations of MG: Lagrange multiplier and singular disformal transformations \cite{golovnev/14}. The first one enables to incorporate the condition (\ref{mgconsist}) at the level of the action through a Lagrange multiplier. Meanwhile, the second formulation highlights the mapping $g \rightarrow \tilde{g}, \varphi$ as a singular disformal transformation in which $\varphi$ corresponds to a new degree of freedom in the gravitational sector \cite{rabochaya/16, sebastiani/17}. 

Here, we are considering MG as a different way to write the effective terms of LQC dynamics as implemented in \cite{langlois/17}, which was the theme of many works in the literature (for example \cite{brahma/18, haro/18, haro/182, haro/183, bodendorfer/18}). The LQC effective Friedmann equation (\ref{h_lqc}) is reproduced by constructing an action whose dynamical variables are $a$, $N$ and $\varphi$. This action must satisfy the requirement of invariance under time reparametrization which, for a flat FRW space-time, can be achieved through the expression \cite{langlois/17}
\begin{equation}
S[a, N, \varphi] = \int dt \left[- \frac{3a\dot{a}^{2}}{8\pi G N}  + a^{3} \frac{\dot{\varphi}^{2}}{2N} + N a^{3}{\cal{L}}\left(a, \frac{\dot{a}}{N}\right) \right].
\label{acao_ming}
\end{equation}

The MG procedure consists in setting $\cal{L}$ like a function of the Hubble parameter ${\cal{F}}(H)$ whose form is defined as an ansatz to obtain the Hamiltonian density,
\begin{equation}
   {\cal{H}} = a^{3} \left[\frac{\pi^{2}_{\varphi}}{2a^{6}} - \frac{8\pi G}{3}\alpha^{2} \sin^{2}\left(\frac{\pi_{a}}{2\alpha a^{2}}\right) \right],
   \label{hden}
\end{equation}
where $N$ continues as a Lagrange multiplier like in LQC \footnote{From now on, the gauge $N=1$ will be assumed for computations regarding the equations of motion, which means we are evolving the system considering the proper time \cite{bojowald/06}.}, $\alpha$ is a constant and
\begin{equation}
   \pi_{\varphi} = \frac{a^{3}}{N}\dot{\varphi} \;\; \mathrm{and} \;\; \pi_{a} = \alpha a^{2} \arcsin{\left( -\frac{3H}{4\pi G \alpha}\right)}
\end{equation}
are the momenta of the scalar field and scale factor, respectively, satisfying 
\begin{equation}
\{ a, \pi_{a} \} = \{ \varphi, \pi_{\varphi}\} = 1.
\label{}
\end{equation}
Note that instead of $p$ and $c$, in MG description, $a$ and $\pi_{a}$ correspond to the pair of non-trivial canonically conjugated variables together with the pair $\varphi$ and $\pi_{\varphi}$ \cite{langlois/17}. Next, a similar procedure to the one previously presented in \ref{slqc} is applied to obtain (\ref{h_lqc}). However, in this case, the energy density and critical energy density are given by
\begin{equation}
\rho = \frac{\pi^{2}_{\varphi}}{2a^{6}} \;\;\; \mathrm{and} \;\;\; \rho_{c} = \frac{8\pi G}{3}\alpha^{2}.
\label{rho}
 \end{equation}
Regarding the equivalence with Effective LQC dynamics, the critical energy density from (\ref{rho}) and (\ref{rhoc}) will be equivalent only if $\alpha$ obeys the relation 
\begin{equation}
\alpha = \frac{3}{8\pi G \gamma \sqrt{\Delta}}.
\label{alpha}
\end{equation}

In \cite{langlois/17}, the generalization for curved space-times is performed by adding a term related to the curvature parameter $k$ in the action

\begin{equation}
\begin{split}
S_{k}[a, N, \varphi] = & \int dt \left[- \frac{3a\dot{a}^{2}}{8\pi G N} \right. \\ 
& + \left. a^{3} \frac{\dot{\varphi}^{2}}{N} + \frac{3Nka}{8\pi G} + N a^{3}{\cal{L}}_{k}\left(a, \frac{\dot{a}}{N}\right) \right],
\label{acao_k}
\end{split}
\end{equation}
and expanding the flat case definition of ${\cal{L}}$ by introducing the curvature dependence within Lagrangian

\begin{equation}
    {\cal{L}}_{k}\left(a, \frac{\dot{a}}{N}\right) = {\cal{F}}(H) -  \frac{3}{8\pi G}V_{k}(a),
    \label{}
\end{equation}
as a potential term $V_{k}(a)$. Consequently, the Hamiltonian density changes and acquires the following form

\begin{equation}
{\cal{H}} = a^{3} \left[ \rho - \rho_{c} \sin^{2}\left(\frac{\pi_{a}}{2\alpha a^{2}}\right) - \frac{3k}{8\pi Ga^{2}}+ \frac{3V_{k}(a)}{8\pi G}\right], 
\label{hamcurv}
\end{equation}
resulting in the modified Friedmann equation for curved space-times \cite{langlois/17}

\begin{equation}
\begin{split}
H^{2} = & \left[\frac{8\pi G}{3}\rho - \frac{k}{a^{2}} + V_{k}(a)\right]  \\
& \times  \left[1 - \frac{1}{\rho_{c}}\left(\rho - \frac{3k}{8\pi Ga^{2}} + \frac{3V_{k}( a)}{8\pi G} \right)  \right]. 
\label{h2_curva}
\end{split}
\end{equation}

It is important to note that the formulation for the curvature mimetic gravity, developed by \cite{langlois/17}, was analyzed by \cite{bodendorfer/18} in terms of whether the curvature is identified with a multiple of the Planck scale. To answer this question the authors analyze if such a relationship can hold in the context of Bianchi I models. The conclusion of the authors is that in the case of Bianchi I spacetime the Hamiltonian for curvature mimetic gravity cannot be interpreted as an effective Hamiltonian arising from loop quantization. However, as emphasized by \cite{cesare/18}, it is unclear if such a limitation may exist for a curvature potential that reproduces the cosmological background dynamics similar to that derived in the group field theory approach to quantum gravity.

Another key point to consider is the instability issue afflicting high derivative mimetic models due to the presence of gradient/ghost. This makes difficult to obtain a stable model capable of reproducing the LQC equations (see, e.g., \cite{firouzjahi/17, zheng/17, gorji/18}). Notwithstanding, it is interesting to evaluate if there are healthy features that can emerge from the MG in order to reproduce the universe dynamics within the scope of the LQC. This is the direction that we intend to discuss the reinterpretation of the MG curvature potential in the next section.

\section{Curvature mimetic gravity: possible scenarios}
\label{discussion}

\subsection{\label{sec:level3}Reinterpreting the potential term of mimetic gravity}

To begin with, from equation (\ref{rho}), the absence of a potential term in the matter density comes from the assumption of a massless scalar field and/or the simplicity argument of defining $V(\varphi) = 0$ that makes easy to perform the quantization process \cite{langlois/17}. On the other hand, if we consider a fundamental field non-minimally coupled to gravity, then it would naturally  be induced a mixing between the kinetic term of the scalar field and the metric field (here represented as a curvature potential in the MG description of LQC). 

In the following computations, we will replace $V_{k}(a)$ by $V_{k}(\varphi)$ in order to make clear our interpretation of the MG curvature potential as a direct response of the matter presence curving the space-time. Moreover, the reverse idea can also be applied, the matter content adapting itself according to the space-time curvature, emphasizing their intrinsic relation. To put this in another way, $V_{k}(\varphi)$ could correspond to the signature of the non-minimal coupling of a fundamental field to the curvature represented by $V_{k}(a)$ at the level of LQC.

This way of describing the primordial universe seems to be a natural interpretation within the scope of LQG since, in this theory, space-time and quantum fields are not distinct components. That is, the space-time we perceive on a large scale is an image generated by quantum fields that `live on themselves'.

Thus, we proceed by reinterpreting equation (\ref{h2_curva}). The strategy applied in \cite{langlois/17} was to introduce the curvature by adding a potential term that only depends on $a$ and $k$ in the gravitational part of the Lagrangian. Meanwhile, the field potential was neglected. Accordingly, we propose to change the field potential from matter sector to gravitational sector as a different way of interpreting the curvature role. First, (\ref{h2_curva}) is rewritten as

\begin{equation}
\begin{split}
 H^{2} = & \frac{1}{3M^{2}_{Pl}} \left[\rho_{\mathrm{kin}} + 3M^{2}_{Pl} \left(V_{k}(\varphi)- \frac{k}{a^{2}}\right) \right] \\   
&  \times \left\{1 - \frac{1}{\rho_{c}}\left[\rho_{\mathrm{kin}} + 3M^{2}_{Pl} \left(V_{k}(\varphi)- \frac{k}{a^{2}}\right) \right]  \right\}, 
\end{split}
\label{nossoh}
\end{equation}
where the kinetic term is
\begin{equation}
\rho_{\mathrm{kin}} =  \frac{\pi^{2}_{\varphi}}{2a^{6}},
\label{rhok}
\end{equation}
and $V_{k}(\varphi)$ is the field potential related with curvature. After, we define an effective energy density as
\begin{equation}
\rho_{\mathrm{eff}} = \rho_{\mathrm{kin}} + 3M^{2}_{Pl} \left[V_{k}(\varphi)- \frac{k}{a^{2}}\right],
\label{reff}
\end{equation}
returning to the primary form of the effective Friedmann equation
\begin{equation}
H^{2} = \frac{\rho_{\mathrm{eff}}}{3M^{2}_{Pl}}  \left(1 - \frac{\rho_{\mathrm{eff}}}{\rho_{c}}\right).
\label{hrhoeff}
\end{equation}

Splitting the kinetic contribution from the potential one could be a strange arrangement at first look. However, this setup enables to treat the field as technically massless from the matter Hamiltonian point of view, once its ``effective mass" contribution could be interpreted as an effect of the non-minimal coupling to gravity. About holonomy corrections that characterize space-time deformations, it will not have any actual difference because they are computed from both gravitational and matter sectors.

The balance among the contribution of the components from $\rho_{\mathrm{eff}}$ during universe evolution needs to be adjusted. This is performed by respecting the changes from LQC energy range  during the primordial universe evolution and the usual requirements for the occurrence of an inflationary period. From (\ref{hamcurv}), by isolating the term with the sine function,

\begin{equation}
\sin^{2}\left(\frac{\pi_{a}}{2\alpha a^{2}}\right) = \frac{1}{\rho_{c}} \left( \rho_{\mathrm{kin}} + \frac{3V_{k}(a)}{8\pi G} - \frac{3k}{8\pi Ga^{2}}\right) = \frac{\rho_{\mathrm{eff}}}{\rho_{c}}, 
\label{}
\end{equation}
we conclude that $\rho_{\mathrm{eff}}$ is the amount to be compared to $\rho_{c}$. During the bounce, (\ref{h_lqc}) is recover for $\rho_{\mathrm{eff}} = \rho_{\mathrm{kin}}$, just as the flat case. However, it could have happened an equilibrium between the two terms related to curvature, $V_{k}(\varphi) = ka^{-2}$, which seems to be a reasonable assumption since the universe motion must stop at the bounce point.

In \cite{mielczarek/17}, the energy range between $(1/2)\rho_{c}$ and $0$ was pointed as the most suitable period to explore the slow-roll approximation. A similar statement can be found in \cite{sadjadi/13} since the onset of the usual inflationary evolution is considered after the universe had achieved $(1/2)\rho_{c}$. Likewise, we are considering the outset of inflationary period around $\rho_{\mathrm{eff}} \simeq (1/2)\rho_{c}$, where we replaced the energy density from LQC by the mimetic form $\rho_{\mathrm{eff}}$. In analogy with the standard slow-roll approach, the kinetic energy will be much smaller than the potential term associated with curvature, reducing (\ref{nossoh}) to
\begin{equation}
H^{2} = \left[V_{k}(\varphi)- \frac{k}{a^{2}}\right]\left\{1 - \frac{3M^{2}_{Pl}}{\rho_{c}} \left[V_{k}(\varphi)- \frac{k}{a^{2}}\right] \right\}. 
\label{nossohi}
\end{equation}
Analyzing the effective energy density $\rho_{\mathrm{eff}} \approx 3M^{2}_{Pl} [V_{k}(\varphi)- ka^{-2}]$, we note how it decreases with expansion, like expected. 

Just as the universe evolves, the increasing scale factor makes the universe geometry becomes flat by diluting any signal of curvature, in agreement with current data from Planck satellite \cite{planck/16}. Therefore, there had been a moment in which the field potential reached an energy range comparable with the kinetic one, sealing the inflationary epoch. At this point, quantum corrections should be negligible, turning (\ref{nossoh}) into 
\begin{equation}
H^{2} \approx  \frac{1}{3M^{2}_{Pl}} \left[\rho_{\mathrm{kin}} + 3M^{2}_{Pl} V_{k}(\varphi) \right].
\label{nossohend}
\end{equation}

Once we obtain equation (\ref{nossohend}) showing the role of the mimetic potential on the universe dynamics as described by the Hubble parameter, we can choose any inflationary field to be described through the mimetic potential $V_{k}(\varphi) - k/a^{2}$. In other words, since inflation under the LQC will occur in the interval $0 \lesssim \rho_{\rm eff}/\rho_{c} \lesssim 1/2$, it is enough to choose a scalar field able of producing inflation and to verify if the energy scales of the inflationary field can be adequately mimicked by LQC in terms of the formulation described by \cite{langlois/17} for the curvature mimetic potential.

Although it is possible to do this analysis with the inflaton, as the scalar field responsible for producing inflation, our choice will fall on the model called Higgs Inflation. There are two reasons for this choice: (1) the only fundamental scalar field with experimental counterpart is the Higgs field, (2) the inflationary version of the Higgs field corresponds to a field not minimally coupled with gravity, a characteristic that seems interesting within the scope of MG. It is this scenario that we will analyze in Sect. \ref{higgs_in}.


\subsection{Curvature potential mimicking the dynamics of Higgs inflation}\label{higgs_in}

The Higgs field playing the role of inflaton, the usual scalar field associated with the standard slow-roll inflation, is an idea that has been discussed since the first inflationary models were developed, as can be seen in \cite{kolb/90}. Notwithstanding, HI describes inflation as a chaotic scenario in which a Higgs field is coupled with the curvature through large values of self-coupling $\lambda$ and non-minimal coupling $\xi$ parameters \cite{bezrukov/08, bezrukov/13, zeynizadeh/15}. Basically, HI reproduces the successful flat potential of slow-roll approximation by coupling a primordial version of the current Higgs field with the Ricci scalar. Besides, $\xi$, $\lambda$, and the relation between them are only determined by cosmological observations \cite{bezrukov/13, postma/14, shaposhnikov/14, calmet/18}.

The HI universe dynamics can be expressed by the action (see \cite{bezrukov/13})

\begin{equation}
S_{J} = \int d^{4}x \sqrt{-g}\left[-\frac{1}{2}(M^{2} + \xi h^{2})R + g_{\mu\nu}\frac{\partial^{\mu} h\partial^{\nu}h}{2} - V(h) \right],
\label{aj}
\end{equation}
where the subscript $J$ means Jordan frame and $V(h)$ is the potential of Higgs field background $h$ constructed like

\begin{equation}
V(h) = \frac{\lambda}{4}(h^{2} - v^{2})^{2},
\label{}
\end{equation}
which is the usual Higgs potential from Standard Model of Particle Physics in the unitary gauge ($2H^{\dagger}H=h^{2}$). Meanwhile, the term $\xi h^{2} R$ corresponds to the non-minimal coupling of the scalar field to curvature.

After electroweak symmetry breaking, the scalar field acquires a non-zero vacuum expectation value (VEV) $v=246 \: {\rm GeV}$ and so $M$ and $\xi$ are then related by $M_{Pl}^2 = M^2 + \xi v^2$. Moreover, as discussed in  \cite{bezrukov/13}, the term $\xi v$ is negligible compared to $M$ for most situations covered by the inflationary Higgs scenario. Therefore, once the parameters $M$ and $M_{Pl}$ differ for the non-zero VEV of $\langle h \rangle = v$, we can consider $M \simeq M_{Pl}$.

Due to the complexity of working with the mixing terms in action (\ref{aj}), the usual procedure is to get rid of the non-minimal coupling to gravity by changing the variables through a conformal transformation from Jordan's frame (the standard one) to Einstein's frame. This transformation has the following form
\begin{equation}
\tilde{g}_{\mu\nu} = \Omega^{2}(h)g_{\mu\nu},
\end{equation}
where
\begin{equation}
\Omega^{2}(h) = \frac{M^{2}+\xi h^{2}}{M^{2}_{Pl}} \approx 1+\frac{\xi h^{2}}{M^{2}_{Pl}},
\label{}
\end{equation}
allowing to write the action in the Einstein frame as
\begin{equation}
\begin{split}
S_{E} = & \int d^{4}x \sqrt{-{\tilde{g}}} \left[-\frac{M_{Pl}^{2}}{2}\tilde{R} \right.\\
& + \left. \left( \frac{\Omega^{2}+6\xi h^{2}/M^{2}_{Pl}}{\Omega^{4}} \right)\tilde{g}_{\mu\nu}\frac{\partial^{\mu} h\partial^{\nu}h}{2} - \frac{V(h)}{\Omega^{4}}  \right]. 
\end{split}
\end{equation}

The conformal transformation produces a non-minimal kinetic term for the Higgs field. Nevertheless, it is possible to obtain a canonically normalized kinetic term through a new field $\chi$ satisfying (see \cite{rubio/09,bezrukov/13})

\begin{equation}
\frac{d\chi}{dh} = \sqrt{\frac{\Omega^{2} + \frac{3}{2}M^{2}_{Pl}({\Omega^{2})'}^{2}}{\Omega^{4}}} = \sqrt{\frac{1 + (\xi +6\xi^{2})h^{2}/M^{2}_{Pl}}{(1 + \xi h^{2}/M^{2}_{Pl})^{2}}},
\label{dxidh}
\end{equation}
here the apostrophe represents the derivative with respect to $h$. It is important to pay attention that $h$ does not change after the conformal transformation, the redefinition (\ref{dxidh}) is just a way to recover the standard form of the slow-roll action,

\begin{equation}
S_{E} = \int d^{4}x \sqrt{-{\tilde{g}}} \left[-\frac{M_{Pl}^{2}}{2}\tilde{R}
+ \tilde{g}_{\mu\nu}\frac{\partial^{\mu} \chi \partial^{\nu} \chi}{2} - V(\chi)  \right].
\end{equation}
Where the potential described in terms of $\chi$, $V(\chi) = V(h)/\Omega^{4}$, leads to a change of the Friedman equation that can be written as

\begin{equation}
H^{2} = \frac{1}{3M_{Pl}^{2}} \frac{V(h)}{(1 + \xi h^{2}/M_{Pl}^{2})^{2}}.
\label{hihub}
\end{equation}
On the other hand, equation (\ref{dxidh}) can be integrated, resulting in (see \cite{rubio/09})

\begin{equation}
\begin{split}
\chi(h) & = \frac{h}{u}\sqrt{1+6\xi}\,\sinh^{-1}\left(\sqrt{1+6\xi}\,u\right) \\
& -\frac{h}{u}\sqrt{6\xi}\,\sinh^{-1}\left(\sqrt{6\xi}\frac{u}{\sqrt{1+u^{2}}}\right),
\end{split}
\label{fullsol}
\end{equation}
with $u=\sqrt{\xi}h/M_{Pl}$.

Since HI is built under the requirement $\xi \gg 1$, if the non-minimal coupling is chosen to be in the range $1 \ll \xi \ll M^{2}_{Pl}/v^{2}$ then equation (\ref{fullsol}) corresponds to the conformal transformation $\Omega^{2} = e^{2\chi/\sqrt{6}M_{Pl}}$. Thus, the potential $V(\chi)$ is given by 
\begin{equation}
V(\chi) = V_{0}\left(1 - e^{-\frac{2\chi}{\sqrt{6}M_{Pl}}}\right)^{2},
\label{pot_einstein}
\end{equation}
with $V_{0} = \lambda M^{4}_{Pl}/4\xi^{2}$. Note that the potential (\ref{pot_einstein}) is exponentially flat for large values of $\chi$, which enables to reproduce an evolution analogously to standard slow-roll inflation in Einstein frame. Therefore, the Friedmann equation presents the form

\begin{equation}
H^{2} \simeq \frac{1}{3M^{2}_{Pl}}V(\chi) \simeq \frac{\lambda M^{2}_{Pl}}{12\xi^{2}},
\label{hi_hubble}
\end{equation}
where it was assumed $\chi \gg \sqrt{3/2}\,M_{Pl}$.

Furthermore, it has been shown (see, e.g., \cite{ber/14,rubio/15,ber/18}) that HI is also in agreement with the most recent estimates obtained through the WMAP and Planck satellites from the cosmic microwave background (CMB) radiation. In particular, CMB normalization requires $\xi \simeq 50000 \sqrt{\lambda}$. Moreover, as discussed by \cite{rubio/15}, for $\xi \sim 10^{3}$, HI is a graceful way to relax to the standard model vacuum.

In \cite{langlois/17}, they establish a link between LQC and MG. Meanwhile, the works \cite{chamseddine/182} and \cite{chamseddine/18} open the possibility to explore MG with the Higgs mechanism. Here, we intend to close this triangle by using MG as the bridge between Effective LQC and HI. We are going to emphasize the intermediary character of the mimetic approach as the one capable of mimicking Effective LQC dynamics besides incorporating the matter-curvature relation from HI. In other words, we intend to answer the following questions: (a) Could curvature mimetic gravity be used to describe the same evolution that HI provides? (b) Are the energy scales of LQC, within MG framework, compatible with the energy scales of HI? (c) If we get affirmative answers to the two previous questions, what form should the curvature potential take?

First of all, considering the equivalence between these two approaches, we can match (\ref{hmax}) with (\ref{hi_hubble}) which will result in
\begin{equation}
\rho_{c} \simeq \frac{\lambda M^{4}_{Pl}}{\xi^{2}}, 
\label{lqc_hi1}
\end{equation}
that is, we can map the behavior of $V_{k}(\varphi) - ka^{-2}$ to mimic the inflationary phase. In terms of energy scale, the inflation occurs in the interval $0 \lesssim \rho_{\rm {eff}}/\rho_{c} \leq 1/2$, thus, (\ref{lqc_hi1}) represents the equality at the onset of inflation.

Due to its structural construction, we are considering MG as a LQC description in the Einstein frame. For $\chi_{\mathrm{end}} \simeq 0.94 M_{Pl}$, the potential $V(\chi)$ represented in (\ref{pot_einstein}) reduces nearly seventy percent of its initial value. Hence, instead of $V(\chi) \approx V_{0}$ we will compute the Friedmann equation with $V(\chi) \approx 0.287 \, V_{0}$ and compare it to 
\begin{equation}
H^{2} \simeq \frac{\rho_{\mathrm{eff}}}{3M^{2}_{Pl}} \approx  \frac{1}{3M^{2}_{Pl}} \left[\rho_{\mathrm{kin}} + 3M^{2}_{Pl} V_{k}(\varphi) \right],
\label{endinf}
\end{equation}
once the quantum gravitational effects should be negligible at this point. As a result, the effective energy density for this period will be determined by
\begin{equation}
\frac{\rho_{\mathrm{eff}}}{3M^{2}_{Pl}} \approx 0.024 \frac{\lambda M^{2}_{Pl}}{\xi^{2}}.
\label{lqc_hi2}
\end{equation}
After, we plug (\ref{lqc_hi1}) in (\ref{lqc_hi2}) and obtain
\begin{equation}
\rho_{\mathrm{eff}} \approx 0.072 \frac{\lambda M^{4}_{Pl}}{\xi^{2}} = 0.072 \rho_{c},
\label{lqc_hi3}
\end{equation}
which is in agreement with the LQC requirement of $\rho_{\mathrm{eff}} \ll \rho_{c}$ in order to recover the classical FRW evolution at the end of the inflationary period.

It is important to note that in \cite{chamseddine/182} was obtained a massive graviton through a Brout-Englert-Higgs (BEH) mechanism in which one of the four scalar fields used was the mimetic gravity field. Notwithstanding this procedure was performed to avoid the appearance of a ghost mode. Note that there is a straight relation between the mimetic field and the Higgs field once BEH is involved. Further, it was also highlighted the strong coupling of the mimetic field with the graviton in scales close to the Planck one, which means a non-minimal coupling between matter and curvature as well as we have been exploring along this work.

On the other hand, the relationships derived above show that it is possible to describe the primordial universe evolution in a unified LQC-HI scenario using as a connection the MG formalism. In this case, the inflationary Higgs field could be `mimicked' from the curvature potential. For that reason, the answers to questions (a) and (b) placed above are `yes' to both. To answer the question (c), we should first establish the relation with HI. Note that to exist a perfect mapping between the curvature potential and $V(\chi)$ during the inflationary phase, the equation

\begin{equation}
\rho^{2}_{\rm eff}-\rho_{\rm eff}\rho_{c}+V(\chi)\rho_{c} = 0
\label{polinon}
\end{equation}
it must be satisfied as a tracking condition. To put it another way, since the relation (\ref{polinon}) comes from the equality between (\ref{hrhoeff}) and (\ref{hi_hubble}), it corresponds to a requirement of the `validity of the mimicry' of the HI scenario through the curvature potential of the MG. Note that $V_{k}(\varphi)$ can be adjusted to allow satisfying equation (\ref{polinon}) throughout an inflationary phase characterized by $V(\chi)$, while controlling $\rho_{\rm eff}$ to be within the usual LQC values.

\begin{figure} 
\begin{minipage}{\columnwidth}
\vspace{2mm}
\centering
{\includegraphics[width= \textwidth]{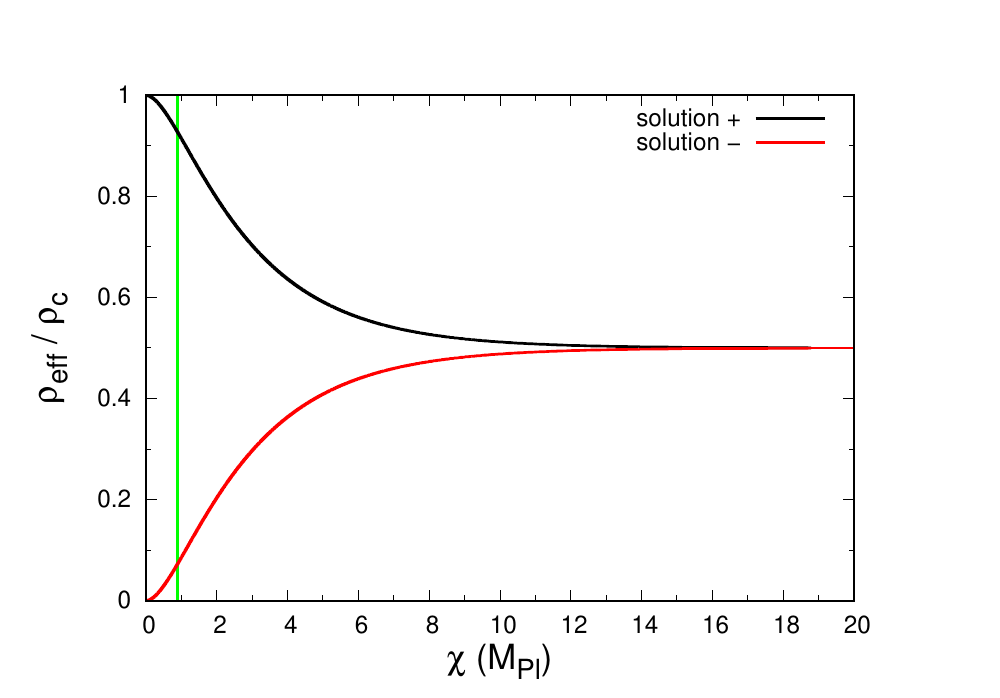}}
\end{minipage}
\caption{The evolution of $\rho_{\rm {eff}} / \rho_{c}$ obtained from MG as a function of the normalized field $\chi$. These solutions satisfy equation (\ref{polinon}). Note that the black line curve does not correspond to a physically consistent solution. The vertical green line indicates the end of the inflationary epoch ($\chi_{\mathrm{end}} \simeq 0.94 M_{Pl}$).}
\label{fig1}
\end{figure}

Because (\ref{polinon}) is a second-degree equation, it has two solutions whose evolution regarding $\chi$ is presented in Figure \ref{fig1}. The physical solution is in red. Meanwhile, the black line describes the non-physical evolution in which the effective potential is growing as the value of $\chi$ decreases. Once the relation $\rho_{\rm eff}/\rho_{c}$ versus $\chi$ is obtained, it is possible to see how the mimetic potential must behave to produce a dynamic similar to that produced by $V(\chi)$.The vertical green line corresponds to the end of inflationary epoch defined at $\chi_{\mathrm{end}} \simeq 0.94 M_{Pl}$.

\begin{figure} 
\begin{minipage}{\columnwidth}
\vspace{2mm}
\centering
{\includegraphics[width= \textwidth]{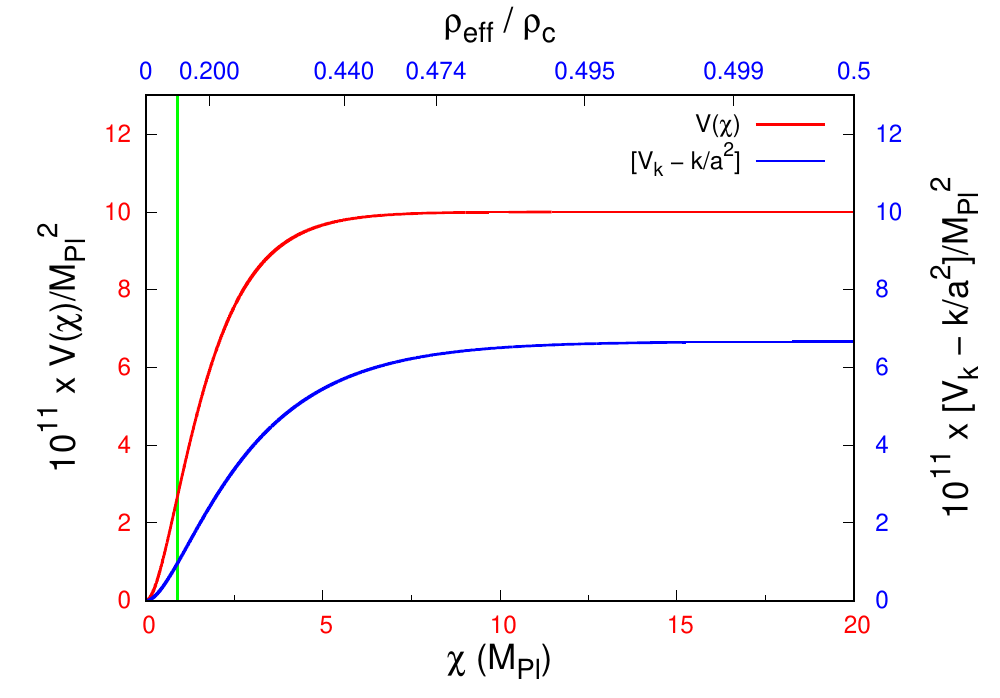}}
\end{minipage}
\caption{The evolution of HI potential and MG curvature potential. The behavior of $V_{k}(\varphi) - ka^{-2}$ mimics the same dynamics, represented by the evolution of the Hubble parameter, as that obtained by the HI scenario. The evolution of potential $V(\chi)$ as a function of the field $\chi$ is presented in the $y1 – x1$ axes (in red). The axes $y2 – x2$ (in blue) show the evolution of $V_{k}(\varphi) - ka^{-2}$ as a function of $\rho_{\rm {eff}} / \rho_{c}$. The vertical green line indicates $\chi_{\mathrm{end}} \simeq 0.94 M_{Pl}$ which represents the end of the inflationary era within the HI approach (see, for example, \cite{bezrukov/08,bezrukov/13}). See that $\chi_{\mathrm{end}} \simeq 0.94 M_{Pl}$ corresponds to $\rho_{\mathrm{eff}}/\rho_{c} \approx 0.072$, value that is in agreement with the LQC requirements for the end of inflation (see, for example, \cite{mielczarek/102}).}
\label{fig2}
\end{figure}

Figure \ref{fig2} exposes the mimetic character of $V_{k}(\varphi) - ka^{-2}$ regarding the behavior of the Hubble parameter given by HI. The evolution of the potential $V(\chi)$ as a function of the Higgs field ($\chi$) is presented in the $y1 - x1$ axes (in red). The behavior that the mimetic potential must have to produce the same $H(t)$ function of the HI scenario is presented in $x2 - y2$ axes (in blue). The dynamic evolution of the universe is the same in both cases so that $V_{k}(\varphi) - ka^{-2}$ can adequately mimic the HI scenario. During the Higgs phase, the scale factor of the universe grows by a number of e-folds $N \simeq 60$ (\cite{ber/18}). Therefore, the term $k/a^{2}$ is diluted and naturally the mimetic potential $V_{k}(\varphi) \rightarrow V(\chi)$ after the end of inflation.

\subsection{Possible relations between matter and curvature in the mimetic gravity representation of loop quantum cosmology}\label{secrelations}

In standard LQC, the relation between matter and curvature is not directly explored, once the sectors are linked but they are not analyzed together as a pair. Since holonomy corrections arise with the area discretization of space-time in area gaps $\Delta$ due to the discrete curvature of Ashtekar connection, the changes affect only the gravitational part of the Hamiltonian. A similar statement can be applied regarding the scalar field whose possible self-interaction may not influence the gravitational sector at all \cite{mielczarek/092}. Therefore, in \cite{mielczarek/092}, they concluded that the evolution during quantum regime is not affected by the introduction of curvature.

Here, we will demonstrate the fundamental role played by the curvature as an essential dynamic element of the MG description of LQC. Considering the definition (\ref{reff}), we are going to show that depending on how the energy density is interpreted, the results can change considerably. Once, despite $\rho_{\mathrm{eff}}$ being the one to follow the LQC energy range evolution, it may or may not be the amount chosen to satisfy the continuity equation (\ref{continuity}). To clarify this, it is important to have in mind that (\ref{continuity}) refers to the matter content. However, we need to specify what amount is playing this role, once assuming only the kinetic term from the start will not have any potential term to drive inflation later. Below, the first case exposed is a straight analogy with the definition (\ref{rho2}), nevertheless, instead of $V(\varphi)$, we have $ V_{k}(\varphi)$ as the matter component in the total energy density. In the second case, we use $\rho_{\mathrm{eff}}$ directly, assuming also the curvature term $ka^{-2}$ related to universe geometry as part of the total energy density. In \ref{app} we provide more details about the validity of the usual continuity equation considering terms related to the curvature.

\subsubsection{Case I: $V_{k}(\varphi)$ as the effective matter potential term in the matter energy density}\label{case_1}

First of all, we define a new variable $\rho$ as

\begin{equation}
\rho= \rho_{\mathrm{kin}} + 3M^{2}_{Pl} V_{k}(\varphi)
\label{rhovk}
\end{equation}
from which the Effective Friedmann equation (\ref{nossoh}) can be written in the form

\begin{equation}
H^{2} = \frac{8\pi G}{3}\left[\rho - 3M^{2}_{Pl} \frac{k}{a^{2}}\right] \left\{1 - \frac{1}{\rho_{c}}\left[\rho - 3M^{2}_{Pl} \frac{k}{a^{2}}\right]  \right\}.
\label{hrho-k}
\end{equation}

Here, $\rho$ represents the total matter energy density. We are considering that the curvature mimetic potential term $3M^{2}_{Pl} V_{k}(\varphi)$ mimics the dynamics of the matter field potential. Therefore, $\rho$ must satisfy the continuity equation (\ref{continuity}). After, we repeat the process presented in Sect. \ref{slqc} for (\ref{hrho-k}) which results in
\begin{equation}
\dot{H} = -\frac{4\pi G}{3}\left[3\rho (1 + w) - 6M^{2}_{Pl} \frac{k}{a^{2}}\right] \left[1 - \frac{2}{\rho_{c}}\left(\rho - 3M^{2}_{Pl} \frac{k}{a^{2}}\right)  \right].
\label{timehrho-k}
\end{equation}
Then we sum (\ref{hrho-k}) with (\ref{timehrho-k}), obtaining the expression 

\begin{equation}
\begin{split}
\dot{H} + H^{2} = & \frac{4\pi G}{3} \left\{\frac{}{}  \rho (-1 - 3w) \right.\\
& - \left. \frac{2}{\rho_{c}}\left(\rho - 3M^{2}_{Pl} \frac{k}{a^{2}}\right)  \left[ \rho (-2 - 3w) + 3M^{2}_{Pl} \frac{k}{a^{2}}\right]  \right\}.
\end{split}
\label{doth2rho-k}
\end{equation}

Finally, from (\ref{reff}) and (\ref{rhovk}) we can also write (\ref{doth2rho-k}) as 

\begin{equation}
\begin{split}
\dot{H} + H^{2} =  & \frac{4\pi G}{3} \left\{\frac{}{} \rho (-1 - 3w) \right.\\
& - \left. \frac{2\rho_{\mathrm{eff}}}{\rho_{c}}\left[ \rho (-2 - 3w) + 3M^{2}_{Pl} \frac{k}{a^{2}}\right]  \right\}.
\end{split}
\label{doth2rhoeff-k}
\end{equation}

Despite the fact that the works \cite{langlois/17} and \cite{mielczarek/092} considered different versions of the LQC Hamiltonian to describe a curved FRW space-time, they could provide similar expressions. Indeed, from Effective Friedmann equation \cite{mielczarek/092}

\begin{equation}
H^{2} := \left[\frac{\dot{p}}{2p}\right]^{2} = \frac{8\pi G}{3} \frac{1}{\rho_{c}} \left[\rho - \rho_{1}(p)\right] \left[\rho_{2}(p) - \rho\right],
\label{frimiel}
\end{equation}
if we replace the approximations $\rho_{1}(p) \approx 3/(8\pi G a^{2})$ and $\rho_{2}(p) \approx \rho_{c} + 3/(8\pi G a^{2})$ by its analog amounts considering (\ref{h2_curva}), $\rho_{1} = 3M^{2}_{Pl} k/a^{2}$ and $\rho_{2} = \rho_{c} + 3M^{2}_{Pl} k/a^{2}$, assuming again (\ref{rhovk}) as matter energy density, we are going to recover (\ref{hrho-k}). Further, the equation (\ref{timehrho-k}) can be seen as a simplified version of its analog expression obtained in \cite{mielczarek/092}, whose expression contains more terms and includes all respective terms from (\ref{timehrho-k}), except for $6M^{2}_{Pl}ka^{-2}$.

Thus, if we consider MG as a skillful tool to deal with the dynamics involved with the LQC, it is possible to re-analyze different scenarios presented in the literature, within the scope of the LQC, obtaining their results by means of a mimetic potential.

As previously stated, the conceptual elegance of LQG does not come only from the way it constructs a quantum theory of gravity from general relativity and quantum mechanics, but also from the simplicity of considering that the universe was initially composed only of quantum fields. However, these fields do not live in space-time, they live on one another so that the space-time that we perceive today it is a blurred and approximate image of one of these fields: the gravitational field. These aspects can be assessed in some way through the mimetic formalism if we consider it as a tool that, through the mimetic potential, allows to explore the quantum effects on the dynamics of the primeval universe.

\subsubsection{Case II: $V_{k}(\varphi) - ka^{-2}$ as part of the total matter energy density}\label{case_2}

In this case, the matter energy density corresponds to equation (\ref{reff}). Therefore, the term $3M^{2}_{Pl}[V_{k}(\varphi) - ka^{-2}]$ is the one that describes the behavior of the matter potential. As we maintain the structure presented in (\ref{reff}), the Effective Friedmann equation is given by (\ref{hrhoeff}). With this in mind, the procedure is similar to the one performed in Sect. \ref{slqc}, however, $\rho_{\mathrm{eff}}$ becomes the variable that needs to obey the continuity equation,
\begin{equation}
\dot{\rho}_{\mathrm{eff}} + 3H (P_{\mathrm{eff}} + \rho_{\mathrm{eff}}) = 0,
\label{continuityeff}
\end{equation}
and also the state equation $P_{\mathrm{eff}} = w \rho_{\mathrm{eff}}$. As a result, the time derivative of $H$ is

\begin{equation}
\dot{H} = -\frac{4\pi G}{3}\rho_{\mathrm{eff}} \left[3 (1 + w) \left(1 - \frac{2\rho _{\mathrm{eff}}}{\rho_{c}}\right)\right].
\label{timehrhok}
\end{equation}

An interesting aspect of this equation is that the super-inflation regime will only depend on $\rho_{\rm eff}$, in particular, within the range $\rho_{c}/2 < \rho_{\rm eff} < \rho_{c}$. See that $\rho_{\rm eff}$ has two components according to equation (\ref{reff}) that are $\rho_{\mathrm{kin}}$ and the curvature mimetic potential. However, in this interval, the super-inflation evolution can happen for any value $w > -1$, with $\dot{H} > 0$. These values cover a wide range of possible fields (or combination of fields), since some Galileon fields ($w>1$), scalar fields without potential ($w=1$), dust-like behavior ($w=0$), until fields with $w \gtrsim -1$, similar to the cosmological constant. Thus, to some extent, the super-inflation phase lies in range $\rho_{c}/2 < \rho_{\rm eff} < \rho_{c}$ for the MG description of LQC as well as the usual LQC.

On the other hand, from the sum of (\ref{hrhoeff}) with (\ref{timehrhok}), we have

\begin{equation}
\dot{H} + H^{2} = \frac{4\pi G}{3}\rho_{\mathrm{eff}} \left[ - 3w -1 + \frac{2\rho_{\mathrm{eff}}}{\rho_{c}} (2 + 3w) \right].
\label{doth2rhoeffk}
\end{equation}

The standard inflation occurs when $\ddot a > 0$ which is equivalent to $\dot{H} + H^{2}>0$. When $\rho_{\rm eff} = \rho_{c}/2$ we have $\dot H = 0$ that set the end of the superinflationary phase (or transition time). In the interval $0 < \rho_{\rm eff} < \rho_{c}/2$ we have $\dot H <0$ and $\dot{H} + H^{2}>0$ and so the universe lies in the normal inflationary phase.

Through equation (\ref{doth2rhoeffk}) it is possible to verify that inflation occurs if the condition 

\begin{equation}
3w +1 < \frac{2\rho_{\mathrm{eff}}}{\rho_{c}} (2 + 3w)
\label{sadjadi_condition}
\end{equation}
is respected. As pointed by \cite{sadjadi/13}, in usual LQC, even fields with non-negative constant state parameters are capable of driving the universe to an inflationary phase, for example, radiation can satisfy the condition given by equation (\ref{sadjadi_condition}). However, in our case, the main regulator of inflation is the mimetic potential embedded in $\rho_{\rm eff}$. That is, the phase called normal inflation is dominated by the mimetic potential term. If it has the form given in Figure \ref{fig2}, then the value of $ w $ will be adjusted to that specific field causing inflation to occur in the usual way.

Another point to note is that (\ref{doth2rhoeff-k}) and (\ref{doth2rhoeffk}) are identical for $k=0$. Further, despite equations (\ref{acelqc}) and (\ref{doth2rhoeffk}) share the same structure, the later contains a richer physics to be explored. With the effective energy density, the original form of the equations of motion related to the flat case is recovered. However, the curvature is intrinsically intricate as a fundamental element to describe the MG version of Effective LQC.

\section{Final remarks}
\label{final}

Recently, \cite{langlois/17} presented an interesting formulation of mimetic gravity under fundamental aspects of loop quantum cosmology. One of their contributions was the introduction of a mimetic curvature potential that preserves all the healthy properties of LQC. In this work, we discuss alternative ways of using this mimetic potential. At first, we demonstrate that the mimetic potential can produce the same dynamics of the so-called Higgs inflation field. The energy scales of HI scenario are properly mimicked and connected to the LQC energy scales during inflation.

In a second moment, we evaluate what should be the form of the effective mimetic potential ($V_{k}(\varphi)-k/a^{2}$) to produce the identical evolution as HI does. Next, we show possible scenarios that may emerge from the relationship between matter and curvature potential within MG framework, results similar to those derived by other authors within the scope of the LQC.

It is important to mention that a recent paper \cite{wan/15} analyzes the cosmology of the primordial universe through the Standard Model of Particle Physics perspective. The authors present a bounce model with the standard Higgs boson whose contraction phase is characterized by an EoS with $w>1$. At the bounce, $w$ reaches large negative values ($w \ll -1$), followed by an inflationary phase for $w = -1$ with nearly 60 e-folds, the same number of e-folds of the HI studied here.

The MG representation offers a simpler alternative scenario for the bounce phase since it preserves all the healthy properties of the usual LQC. On the other hand, the original formulation of MG provided an interesting alternative to evaluate the dark matter content of the universe, since the dark components are treated as geometric effects \cite{sebastiani/17}.

Our main motivation for exploring MG's curvature potential is that it does not introduce major modifications to the usual LQC structure, as discussed above. Just like the mimetic dark matter-gravity model can be considered a minimal extension of GR \cite{rabochaya/16}. Moreover, within certain limits, it reproduces the formulation studied by \cite{sadjadi/13} ($k = 0$) and the scenario presented in \cite{mielczarek/092} ($k = 1$) depending on how the mimetic potential is considered in the dynamical equations. This is in agreement with the statement already related to usual Mimetic Gravity that discourses about obtaining different cosmological solutions through the suitable choice of the mimetic potential (see \cite{sebastiani/17} and \cite{chamseddine/14} for more details).

In a certain respect, the effective mimetic potential $V_{k}(\varphi)-k/a^{2}$ can be grouped in different ways into the LQC equation, somewhat similar to the cosmological constant originally introduced by Einstein on the geometric side of the general relativity field equations and reinterpreted in the 1980s as a fluid (and therefore moved to the side of the energy-momentum tensor) able to produce the current acceleration of the universe.


\begin{acknowledgements}
We would like to thank the referee for his/her useful comments that improved this work. This study was financed in part by the Coordena\c{c}\~ao de Aperfei\c{c}oamento de Pessoal de N\'ivel Superior (CAPES) - Finance code 001. O.D.M. thanks CNPq for partial financial support (grant 303350/2015-6).
\end{acknowledgements}


\appendix

\section{Energy density conservation}
\label{app}

In Sects. \ref{case_1} and \ref{case_2}, we state that the usual conservation equation given by (\ref{continuity}) holds with the presence of mimetic curvature potential. In order to justify this assertion, let us define a new variable $V_{\mathrm{eff}}$ named MG effective curvature potential in the form

\begin{equation}
V_{\mathrm{eff}} = 3M^{2}_{Pl} [V_{k}(\varphi)- ka^{-2}],
\label{veff}
\end{equation}
where $V_{\mathrm{eff}}$ represents the mimetic potential for the case studied in Sect. \ref{case_2}. In this way, we can rewrite (\ref{reff}) as

\begin{equation}
\rho_{\mathrm{eff}} = \rho_{\mathrm{kin}} + V_{\mathrm{eff}}.
\label{rveff}
\end{equation}
Next, we plug (\ref{rveff}) into (\ref{continuityeff}) and obtain the expression

\begin{equation}
\dot{\rho}_{\mathrm{kin}} + 3H (1+ w) \rho_{\mathrm{kin}} + [\dot{V}_{\mathrm{eff}} + 3H (1+ w)  V_{\mathrm{eff}}] = 0.
\label{pot_a3}
\end{equation}
The first possibility for the left-hand side of equation (\ref{pot_a3}) to be equal to zero, preserving equality, is that the variables $\rho_{\mathrm{kin}}$ and ${V}_{\mathrm{eff}}$ `work together' like in the usual energy density for an ordinary scalar field since it obeys the continuity equation. From Figure \ref{fig2}, we show that $V_{\mathrm{eff}}$ reproduces the behavior of matter potential $V(\chi)$. Therefore, it is reasonable to consider that ${V}_{\mathrm{eff}}$ continues to reflect this behavior along the field primordial evolution.

The second possibility for satisfying (\ref{pot_a3}) is to consider that

\begin{equation}
\dot{\rho}_{\mathrm{kin}} + 3H (1 + w)\rho_{\mathrm{kin}} = 0,
\label{continuity_original}
\end{equation}
which implies the following condition for the evolution of the mimetic potential,

\begin{equation}
\dot{V}_{\mathrm{eff}} + 3H (1+ w)  V_{\mathrm{eff}} = 0.
\label{continuity_veff}
\end{equation}
Using (\ref{veff}) in (\ref{continuity_veff}), it is possible to obtain 

\begin{equation}
\dot{V}_{k}(\varphi) + 3H (1 + w) V_{k}(\varphi) -  (1 + 3w) H \frac{k}{a^{2}} = 0.
\label{dotvkk}
\end{equation}
Note that if the equation (\ref{dotvkk}) is satisfied, then the continuity equation expressed in (\ref{continuityeff}) will be preserved.

For the case in Sect. \ref{case_1} the same analysis can be applied, so that the continuity equation (\ref{continuity}) is preserved for the same condition expressed in (\ref{continuity_veff}). The only difference is that we have $V_{\mathrm{eff}} = 3M^{2}_{Pl} V_{k}(\varphi)$, since the curvature from the term $ka^{-2}$ is considered separately. Thus, the evolution equation for the mimetic potential becomes

\begin{equation}
\dot{V}_{k}(\varphi) + 3H (1 + w) V_{k}(\varphi) = 0.
\label{dotvkk_2}
\end{equation}

With conditions (\ref{dotvkk}) and (\ref{dotvkk_2}) satisfied respectively for cases Sects. \ref{case_2} and \ref{case_1}, we preserve the usual continuity equation. For this second possibility, observe that if $w = -1/3$, then the continuity equation and the acceleration equation for both Case I and Case II will be identical without considering $k = 0$. Moreover, $w = -1/3$ is in agreement with LQC requirements $\dot{H} < 0$, $\dot{H} + H^{2} >0$ and (\ref{sadjadi_condition}) for the inflationary phase, as it was mentioned in Sect. \ref{case_2}.

It is important to note, as discussed by \cite{yoko/14}, that the slow-roll phase during inflation is satisfied for $w < - 1/3$. Additionally, $w$ should have values $\gtrapprox \, { - 1/3}$ for inflation to come to an end. At this point, there follows a reheating phase where $w$ changes from $-1/3$ to some value within the range $[0,1/3]$ if the inflationary phase was governed by the HI mechanism (see \cite{cook/15}).

Thus, if $w$ varies with the scale factor, the conditions expressed in (\ref{dotvkk}) and (\ref{dotvkk_2}) should be changed in accordance with the evolution of $w$. This is a feature that the introduction of mimetic curvature potential can bring to the usual formulation of LQC.


\end{document}